\documentclass[acmsmall,screen,nonacm]{acmart}

\usepackage{amsmath,amsfonts,amsthm}
\usepackage{xspace}
\usepackage{booktabs}
\usepackage{multirow}
\usepackage{subcaption}
\usepackage{enumitem}
\usepackage{microtype}
\usepackage{needspace}
\usepackage{placeins}
\usepackage{flafter}
\usepackage{algorithm}
\usepackage{algpseudocode}
\usepackage{tikz}
\usetikzlibrary{arrows.meta,positioning,fit,backgrounds,calc,shapes.geometric}
\microtypesetup{protrusion=true,expansion=true}
\newcommand{\TOSEMVERSION}{}

\AtBeginDocument{%
  }

\setcopyright{none}
\acmDOI{}

\newcommand{\kw}[1]{{\ensuremath {\mathsf{#1}}}\xspace}

\newcommand{\delete}[1]{}
\newcommand{\notation}[1]{}

\definecolor{c1}{RGB}{42,99,172} %
\definecolor{c2}{RGB}{255,88,93}
\definecolor{c3}{RGB}{255,181,73}
\definecolor{c4}{RGB}{119,71,64} %
\definecolor{c5}{RGB}{228,123,121} %
\definecolor{c6}{RGB}{208,167,39} %
\definecolor{c7}{RGB}{0,51,153}
\definecolor{c8}{RGB}{56,140,139} 
\definecolor{c9}{RGB}{0,0,0} 

\newcommand{\reffig}[1]{Figure~\ref{fig:#1}}
\newcommand{\subreffig}[2]{Figure~\ref{fig:#1}(\subref*{fig:#2})}
\newcommand{\refsec}[1]{Section~\ref{sec:#1}}
\newcommand{\reftable}[1]{Table~\ref{tab:#1}}
\newcommand{\refalg}[1]{Algorithm~\ref{alg:#1}}

\newcommand{\sys}{\kw{GraphWeave}}

\title{Uncovering Systematic Blind Spots in Graph Database Testing through Integration and Process-Level Analysis}

\author{Tinghan Li}
\orcid{0009-0001-6490-6210}
\email{225045006@link.cuhk.edu.cn}
\affiliation{%
  \department{School of Data Science}
  \institution{The Chinese University of Hong Kong, Shenzhen}
  \city{Shenzhen}
  \country{China}}

\author{Aoyang Fang}
\email{aoyangfang@link.cuhk.edu.cn}
\affiliation{%
  \institution{The Chinese University of Hong Kong, Shenzhen}
  \city{Shenzhen}
  \country{China}}

\author{Longbin Lai}
\email{longbin.lai@gmail.com}
\affiliation{%
  \institution{Alibaba Group}
  \city{Hangzhou}
  \country{China}}

\author{Zifan Xie}
\email{xzff@cqu.edu.cn}
\affiliation{%
  \institution{Chongqing University}
  \city{Chongqing}
  \country{China}}

\author{Hongbo Zhao}
\email{hongbozhao72@gmail.com}
\affiliation{%
  \institution{Beijing University of Posts and Telecommunications}
  \city{Beijing}
  \country{China}}

\author{Yu Shao}
\email{yu.shao@alibaba-inc.com}
\affiliation{%
  \institution{Alibaba Group}
  \city{Hangzhou}
  \country{China}}

\author{Ying Zhang}
\email{yingz.au@gmail.com}
\affiliation{%
  \department{Computer Science and Engineering}
  \institution{Zhejiang Gongshang University}
  \city{Hangzhou}
  \country{China}}

\author{Pinjia He}
\email{hepinjia@cuhk.edu.cn}
\affiliation{%
  \department{School of Data Science}
  \institution{The Chinese University of Hong Kong, Shenzhen}
  \city{Shenzhen}
  \country{China}}

\renewcommand{\shortauthors}{Tinghan Li et al.}


\begin{document}


\begin{abstract}
Graph databases are becoming foundational infrastructure for connected data, and a growing body of automated testing techniques has uncovered many bugs across mature graph database systems.
Each technique checks a different aspect of correctness and exposes a different class of defect.
Combining them is the natural next step, but each technique is judged by the real bugs it finds, not by useless reports or tests that can never fail, and that is what decides whether a developer can afford to fix the problems.

To identify the remaining gaps, we integrated test capabilities from 8 testing methods and conducted an empirical study of their behavior, during which we found two blind spots.
First, 12 hours of testing produce thousands of bug reports, far more than the number of genuine defects a mature database is likely to contain.
Most reports are false positives caused by dialect or pattern-matching differences, and many of the remaining reports are duplicates.
Second, analysis of process-level data reveals that many equivalence-based metamorphic test cases lose their ability to detect defects because of the optimization mechanism.

Based on these findings, we summarize and realize these guidelines in \sys, an integrated testing platform for graph databases.
This includes a multi-agent system that analyzes the flood of bug reports to produce gating rules for issue filtering and adapter rules for dialect adaptation, and a plan-aware triage routes metamorphic pairs by plan, skipping those that share a plan on the same graph state and therefore cannot detect defects.
We apply \sys to NeuG, a recently open-sourced industrial graph database that has not been covered by prior graph database testing work.
With modest manual effort, \sys uncovers 19 previously unknown bugs, all of which have since been confirmed.

\end{abstract}

\maketitle
\noindent\textit{Author preprint. Submitted to ACM Transactions on Software Engineering and Methodology; not yet peer reviewed or accepted.}\par\medskip

\section{Introduction}
\label{sec:introduction}


Graph databases are widely used for managing highly connected data, including social networks, recommendation systems,
fraud detection, and knowledge graphs~\cite{robinson2015graph,angles2008survey,hogan2021knowledge}.
Reflecting this demand, new systems continue to appear: as of this writing, the CMU Database of Databases~\cite{dbdb} lists 26 graph database systems first released since 2021.
Each of these new systems must be tested for correctness before it can be relied upon.

Driven by this demand, researchers have developed a range of automated testing techniques for graph databases.
These techniques rely on different test oracles, which are mechanisms for determining whether a test has passed or failed.
The kinds of correctness they check can differ substantially: some compare the results that independent systems return for the same query~\cite{hua2023gdsmith},
while others check, within a single system, whether write operations leave the graph in the expected state~\cite{liu2024testing}.
Different methods therefore expose different classes of defects~\cite{liu2024testing,mang2025finding}.
Combining them is the natural next step, yet no prior work has run these methods together,
and their combined behavior remains unknown.
Each technique is judged by the real bugs it finds, not by useless reports or tests that can never fail, and that is what decides whether a developer can afford to fix the problems.
To identify the remaining gaps, we integrate eight methods~\cite{zhuang2023testing,mang2025finding,kamm2023testing,jiang2024detecting,liu2024testing,mang2024testing,wust2024dinkel,hua2023gdsmith} behind a shared interface, record process-level data for each test, and study what these methods report and why.
The study reveals two blind spots of existing testing methods; \reffig{intro} previews the three mechanisms behind them.

\newcommand{\introvscale}{1.0}
\begin{figure*}[t!]
  \centering
  \begin{subfigure}[t]{0.305\textwidth}
    \centering
    \raisebox{-\height}{\scalebox{1}[\introvscale]{\includegraphics[width=\linewidth]{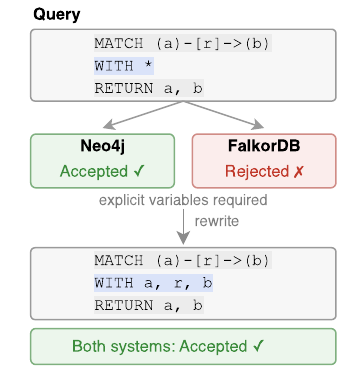}}}
    \caption{Dialect differences}
    \label{fig:intro-dialect}
  \end{subfigure}
  \hfill
  \begin{subfigure}[t]{0.338\textwidth}
    \centering
    \raisebox{-\height}{\scalebox{1}[\introvscale]{\includegraphics[width=\linewidth]{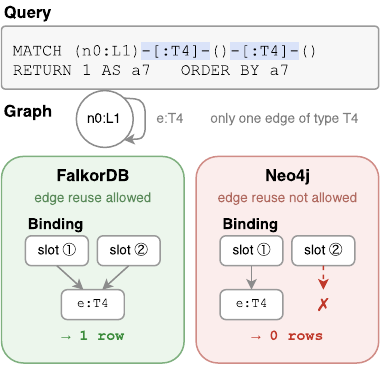}}}
    \caption{Semantic differences}
    \label{fig:intro-semantic}
  \end{subfigure}
  \hfill
  \begin{subfigure}[t]{0.316\textwidth}
    \centering
    \raisebox{-\height}{\scalebox{1}[\introvscale]{\includegraphics[width=\linewidth]{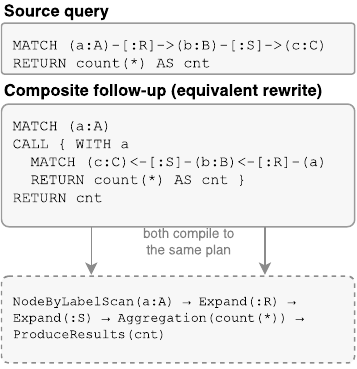}}}
    \caption{Plan collapse}
    \label{fig:intro-plan}
  \end{subfigure}
  \caption{Three fundamental challenges for graph database testing with existing methods:
    (a)~\textit{dialect differences}, (b)~\textit{semantic differences}, and (c)~\textit{plan collapse}.}
  \Description{A three-panel overview of graph database testing challenges. The first panel shows that dialect differences can cause the same logical query to fail across systems. The second panel shows that semantic differences can make two systems return different results for the same pattern. The third panel shows that optimizer plan collapse can remove the behavioral difference needed by metamorphic testing.}
  \label{fig:intro}
\end{figure*}

\paragraph*{Blind spot 1: Most raw reports are false positives or duplicates}
A twelve-hour run of the integrated methods on Neo4j, Memgraph~\cite{Memgraph}, and FalkorDB, three mature graph databases,
produced more than four thousand raw reports,
a volume far exceeding the number of genuine defects these mature systems are likely to contain.
After inspecting these reports with experts,
we found that most reports are false positives.
Although false positives have several sources, two classes are especially important for integrated testing.
The first is \emph{dialect incompatibility}, which arises when graph database systems use different syntax to express the same query semantic.
As \subreffig{intro}{intro-dialect} shows, test cases originally designed for one system are therefore rejected by another and recorded as raw reports.
Prior methods adapt to a target system's dialect through substantial manual effort.
The second arises from differences in the \emph{semantic constraints} of pattern matching~\cite{angles2017foundations}, the core operation of graph queries.
Systems may make different but permissible choices for unspecified semantics, such as whether an edge may be bound by multiple variables within one match.
An oracle built around one choice then produces false positives on a system that makes another.
\subreffig{intro}{intro-semantic} shows one such case.
On a graph with one node and one self-loop, the same pattern returns one row on FalkorDB~\cite{FalkorDB} but zero on Neo4j~\cite{Neo4j},
because FalkorDB permits this reuse of an edge while Neo4j does not.
Such a false positive can be hard to distinguish from a genuine defect without graph-database domain knowledge.
After false positives are filtered,
the remaining reports still number in the hundreds.
Many are redundant because multiple test cases trigger the same underlying bug.
In one campaign, 278 reports remained after 249 false positives were removed, but all 278 pointed to the same bug.
Separating the few real defects from this noise still demands substantial manual effort.

\paragraph*{Blind spot 2: Many metamorphic tests cannot detect defects}
Methods based on equivalence metamorphic relations generate a semantics-preserving variant of each query and check whether both return the same result.
Query optimization mechanism, however, is designed to erase exactly such surface differences and often compiles both queries to the same plan, as shown in \subreffig{intro}{intro-plan}.
When both queries are executed against the same graph state and compile to the same plan, their results are identical by construction, so such a test cannot reveal a defect and only consumes testing resources.
Worse, if that shared plan contains a bug, both queries return the same incorrect result, and the oracle remains silent.

The problems these blind spots cause recur across testing methods and point to concrete guidelines for integrated testing.
We realize these guidelines through the following designs.

\paragraph*{Postprocessing pipeline for noise reduction}
The study shows that in semantic regions where systems may validly differ,
cross-system disagreement alone is weak bug evidence,
while intra-system contradiction across equivalent forms provides stronger evidence of a real defect.
This finding guides us to preserve and analyze process-level data, so that self-contradiction can be detected
and false positives attributed to their underlying semantic assumptions.
We realize this with a postprocessing pipeline.
In this pipeline, reports are clustered by root cause,
and representative cases are analyzed by a multi-agent LLM system with skills that encode findings from the study.
Each confirmed disposition triggers a corresponding feedback action.
The pipeline updates the target adapter for dialect false positives
and registers gate rules for confirmed bugs and duplicates to suppress repeated analysis.

\paragraph*{Plan-aware triage for skipping powerless pairs}
The study shows that 5\% to 34\% of comparable equivalence-based metamorphic pairs compile to the same execution plan on the same graph state, at which point their results are identical by construction and the oracle has no power.
This finding guides us to identify same-plan pairs before result execution.
We realize this with a triage scheme that collects and compares plan hashes.
Same-plan pairs are skipped before result execution.
Pairs with different plans but identical results may indicate missed optimization opportunities or performance issues.
Pairs with different plans and different results provide potential evidence of a logic bug and are passed to the postprocessing pipeline.

\sys combines the initial integration framework with these two designs.
The resulting end-to-end platform supports test construction, dialect adaptation, report postprocessing, and process-level analysis.
Together, these components make testing new graph databases practical with modest computational and human resources.
We apply \sys to NeuG~\cite{NeuG}, a recently open-sourced industrial graph database not covered by prior graph database testing work, and uncover 19 previously unknown bugs, all since confirmed.

\FloatBarrier
We make the following contributions.

\begin{itemize}
\item \textbf{An empirical study of integrated graph database testing.}
We integrate eight methods behind one interface and conduct an empirical study of their behavior,
uncovering two blind spots of existing testing methods.

\item \textbf{Guidelines for integrated graph database testing.}
We summarize these findings as guidelines for separating false positives from real bugs and triaging powerless metamorphic pairs.

\item \textbf{Realizing the guidelines in \sys.}
We build \sys to realize these guidelines and apply it to NeuG,
uncovering 19 previously unknown bugs, all since confirmed.
The distribution of these bugs shows the value of integration, as no single method finds more than half.
\end{itemize}

\section{Background}
\label{sec:background}

\subsection{Testing Methods for Graph Databases}
\label{sec:bg-testing-methods}
The core difficulty of testing is checking the result for an arbitrary query on an arbitrary database state.
Prior work resolves this challenge by proposing \emph{test oracles},
each of which is a consistency relation the result of the query should satisfy.
Existing approaches have proposed oracles to find hundreds of defects in real-world graph databases,
and we classify their test capabilities into three categories~\cite{hua2023gdsmith,kamm2023testing,jiang2024detecting,mang2024testing,zhuang2023testing,wust2024dinkel,liu2024testing,mang2025finding}.

\paragraph*{Differential Testing} Differential testing builds on an assumption:
if the same queries are executed against multiple graph database systems on the same database state,
the results should be the same~\cite{hua2023gdsmith}. Therefore, the method checks correctness by comparing results across systems without knowing the correct answer.
However, differential testing has two limitations.
First, false negatives arise when multiple systems share the same defect, since the oracle cannot detect a shared failure.
Second, false positives arise even on features that all systems support. The language standard leaves parts of the semantics under-specified, so different systems may resolve them in different yet legitimate ways. The oracle then flags this divergence as a defect although no system is actually wrong.

\paragraph*{Statement Metamorphic Testing}
Motivated by the limitations of differential testing,
researchers have designed many metamorphic testing methods.
Statement metamorphic testing transforms the query and checks that the result of the transformed query satisfies an expected metamorphic relation with that of the original~\cite{chen2018metamorphic}.
A common case is equivalence transformation, which rewrites a query into a semantically equivalent one,
such as reordering clauses or introducing redundant conditions, and checks that the two results are identical~\cite{mang2024testing}.
Other methods instead adopt transformations that are not equivalence-preserving.
For instance, TLP partitioning~\cite{rigger2020finding} splits a query into the \texttt{where true}, \texttt{where false}, and \texttt{where null} branches,
whose results differ from the original but together reconstruct it through their union.
Some statement transformations are graph-native,
deriving related queries from structures such as nodes, edges, and paths.
Gamera~\cite{zhuang2023testing} includes such query-derived capabilities through graph-aware metamorphic relations.

\paragraph*{Graph-Derived Testing} Graph-derived testing constructs oracles by operating on the test graph itself rather than on the query~\cite{zhuang2023testing,mang2025finding}.
The underlying observation is that controlled transformations of the graph induce predictable changes in query results;
a violation of the predicted relationship signals a defect.
GSlicer~\cite{mang2025finding} instantiates this idea for graph databases: it decomposes the test graph into exhaustive,
non-overlapping subgraphs and checks that querying each partition and aggregating the results reconstructs the result of querying the full graph.
Gamera~\cite{zhuang2023testing} also includes graph-derived capabilities alongside its query-derived ones.

\subsection{Graph Database Semantic Constraints}
\label{sec:bg-semantic-constraints}

Graph pattern matching maps a structural template onto a property graph,
governed by \emph{semantic constraints} that decide which mappings are valid and how many result rows each produces.
Theory grounds these rules in the choice between isomorphism-based and homomorphism-based semantics~\cite{angles2017foundations};
the Cypher specification adopts the \emph{no-repeated-edge} variant, where edge variables map one-to-one but node variables may repeat~\cite{francis2018cypher}.
Its formal semantics covers only the core read-only fragment, leaving these rules underspecified~\cite{francis2018formal},
and GQL~\cite{GQL2024,deutsch2022graph} designates path semantics as user-selectable options.

We make this design space precise with three constraints on a match within a uniqueness scope $U$.
The match exposes a \emph{visible relationship sequence} $\rho=\langle e_1,\dots,e_k\rangle$ of relationship ids,
which may arise from several \emph{traversal witnesses} $\omega$ with visible projection $\mathsf{vis}(\omega)$, such as the two directions of an undirected edge.
Its \emph{multiplicity} under bag semantics counts these witnesses,
\[
  m(\rho) \;=\; \bigl\lvert \{\, \omega \mid \mathsf{vis}(\omega)=\rho \,\} \bigr\rvert .
\]
A system's profile then fixes three choices: \emph{legality}, whether a sequence with $e_i=e_j$ for some $i\ne j$ is admissible within $U$;
\emph{multiplicity}, the value $m(\rho)$;
and the \emph{reset boundary}, the syntactic point at which $U$ restarts, which is meaningful only when legality forbids reuse.
For oracles that rely on result consistency, a profile mismatched to the target system yields systematic false positives rather than real defects.

\subsection{Query Optimization in Graph Databases}
\label{sec:bg-query-optimization}

The query optimizer is a core component of graph database systems:
before execution, it compiles the user-submitted query into a physical execution plan of minimal estimated cost~\cite{selinger1979access,chaudhuri1998overview}.
The literature categorizes graph query optimization into two complementary classes~\cite{lyu2025modular}.
\emph{Rule-based rewriting} applies heuristic, semantics-preserving transformations that restructure a query into a more compact equivalent form,
such as pushing selections and projections toward the scan, trimming unused fields, fusing adjacent pattern segments, and eliminating redundant or implied pattern constraints~\cite{pirahesh1992extensible,lyu2025modular}.
\emph{Cost-based optimization} then searches the space of physical plans induced by these rewrites and selects the cheapest under statistical cost estimates~\cite{selinger1979access,chaudhuri1998overview}.
For graph pattern matching this search is dominated by join-order selection over the pattern's relationships~\cite{lyu2025modular}.
The optimizer estimates intermediate-result cardinalities for subpatterns, increasingly relying on high-order summaries such as short-path and small-pattern statistics~\cite{park2020g,wang2025path,lyu2025modular},
and chooses among binary-join and worst-case-optimal-join strategies for expanding the pattern~\cite{selinger1979access,chaudhuri1998overview,ngo2018worst,veldhuizen2014leapfrog,lyu2025modular}.
Although the two classes pursue different objectives,
with rule-based rewriting removing structural redundancy
and cost-based optimization minimizing estimated cost,
both converge structurally distinct queries onto a smaller space of canonical plans.

These optimizations carry a critical implication for testing.
Metamorphic oracles rely on the structural difference between a reference query and its follow-up to induce distinguishable execution behavior~\cite{chen2018metamorphic,mang2024testing}.
Yet the optimizer's objective is precisely to eliminate such differences.
When two queries are executed against the same graph state and reduce to the same execution plan, their results agree regardless of whether the system under test contains a defect, and the oracle's consistency check passes unconditionally~\cite{pirahesh1992extensible,lyu2025modular}.
This plan convergence renders affected metamorphic tests incapable of detecting execution-level defects.

\section{Method Integration Framework}
\label{sec:system}

\subsection{A Unified Testing Model}
\label{sec:system-model}

Existing graph database testing methods differ substantially in their construction logic: 
graph-derived testing applies controlled transformations to the test graph and checks the predicted relationship between the results; 
statement metamorphic testing transforms the query and checks a metamorphic relation between the original and transformed results, for example by rewriting a query into a semantically equivalent form; 
differential testing executes a fixed query across multiple database systems and compares their results. 

Despite these surface differences, a single round of testing always follows the same structure:
prepare one or more database states, execute operations on a target system, collect the results,
and check a predefined relation over them.
We formalize this with four elements.
A \textbf{database state} $S$ is a snapshot of a graph database: its nodes, edges, labels, and properties.
An \textbf{operation} $Q$ is a Cypher statement.
Given a state $S$, an operation $Q$, and a concrete database instance $t$ as the execution target,
the \textbf{observation} is everything the oracle inspects, such as the returned rows and the execution plan:
\[
  O \;=\; \mathit{execute}(S, Q, t).
\]
A \textbf{correctness relation} $R(O_1, \ldots, O_n)$ is a predicate over observations; a failing relation constitutes a bug report.

Within this model, the three families differ in exactly one way.
Each test capability varies a single argument of $\mathit{execute}(S, Q, t)$ while fixing the other two, as shown in \refalg{testing-families}.

\begin{algorithm}[!htbp]
\caption{Unified execution skeleton. Each testing family varies one axis of \textit{execute} and fixes the other two.}
\label{alg:testing-families}
\ifdefined\TOSEMVERSION\footnotesize\fi
\begin{algorithmic}[1]
\Require seed inputs $(S, Q, t)$; relation $R$; a derivation $\delta$ expanding one axis:
\Statex \hspace{1.5em}\textsc{GraphDerived}: $\delta = f$ over $S$
\Statex \hspace{1.5em}\textsc{StatementMetamorphic}: $\delta = g$ over $Q$
\Statex \hspace{1.5em}\textsc{Differential}: $\delta = \mathcal{T}$ over $t$
\Ensure a bug report iff $R$ fails over the collected observations
\State $\langle a_0, \ldots, a_n \rangle \gets \delta(S, Q, t)$ \Comment{expand the chosen axis}
\For{$i \gets 0, 1, \ldots, n$}
  \State $x_i \gets (S, Q, t)$, chosen axis set to $a_i$
  \State $O_i \gets \mathit{execute}(x_i)$
  \State $O_i.\mathit{result} \gets \mathit{normalize}(O_i.\mathit{result})$
\EndFor
\State \Return $R(O_0, O_1, \ldots, O_n)$
\end{algorithmic}
\end{algorithm}

\textsc{GraphDerived} varies the state $S$ by deriving a family of states through $f$, \textsc{StatementMetamorphic} varies the operation $Q$ by deriving variant queries through $g$, and \textsc{Differential} varies the target $t$ over a set of systems $\mathcal{T}$.
Everything else, which covers initialization, execution, and result normalization, is identical across families; only how an input is constructed and which relation $R$ is checked differ.
The test capabilities we integrate from eight methods~\cite{zhuang2023testing,mang2025finding,kamm2023testing,jiang2024detecting,liu2024testing,mang2024testing,wust2024dinkel,hua2023gdsmith} reduce to these three families under this interface.
\reftable{eval-setup}(\subref*{tab:eval-methods}) lists the integrated methods by family and publication year.

\subsection{Integrating Methods Behind the Interface}
\label{sec:system-caseplan}

The model above is what makes integration tractable.
Because all test capabilities share the same skeleton and differ only in $f$, $g$, and $R$, \sys can run them in one environment.
Each capability is specified by its construction logic and its correctness relation; a shared engine then performs the uniform execute-and-normalize steps, and a per-target adapter isolates all system-specific logic.
Consequently, integrating a new capability reduces to supplying its construction logic and relation, and supporting a new database reduces to adding one adapter, which requires no change to the engine or to any existing capability.

Crucially, the same shared execution records \emph{process-level data} that individual methods discard: how each state and query was derived and which semantic preconditions each relation relies on.
This trace is what \S\ref{sec:whitebox} and \S\ref{sec:evaluation} build on, because it lets a false positive be attributed to the specific assumption it violates rather than an unexplained mismatch, which is impossible when only the final verdict is kept.

\section{Studying Testing Methods through Process-Level Analysis}
\label{sec:study}

Having integrated eight existing methods behind the shared interface of \S\ref{sec:system},
we now use this platform to study the integrated methods themselves.
Because every campaign records process-level data such as execution plans alongside each verdict,
we can look beyond whether an oracle fired and ask \emph{why} it fired,
attributing each outcome to the concrete construction trace and semantic assumption behind it.
This section reports what that process-level view reveals about existing testing methods.

\subsection{Study Setup}
\label{sec:study-setup}
\label{sec:system-prestudy}

We instantiate our testing framework with the eight methods listed in \reftable{eval-setup}(\subref*{tab:eval-methods}),
and run all of them against three mature, widely deployed graph databases, which in this study are Neo4j, Memgraph~\cite{Memgraph}, and FalkorDB, as shown in \reftable{eval-setup}(\subref*{tab:eval-systems}).
The popularity indicators in \reftable{eval-setup}(\subref*{tab:eval-systems}) are from June 2026: ranks follow DB-Engines~\cite{DBEngines}, and star counts are from GitHub.
All systems run with their default query optimizers enabled and differ in concrete query semantics relevant to oracle validity.
Each $(\textit{method}, \textit{system})$ campaign runs in an isolated container against a separate database instance for a fixed 12-hour budget under identical resource and timeout policies,
and every run records process-level data such as execution plans and execution times.
During integration, we tag each test capability with its originating method, so every report can be attributed to the method that produced it.
We distinguish three levels of outcome used throughout the study: a \emph{raw report} is any oracle-triggered candidate, a \emph{false positive} is a raw report that does not correspond to a real defect, and a \emph{unique bug} is a root-cause-level defect obtained after deduplication.

\begin{table}[H]
  \caption{Testing methods and database systems used in the study.}
  \label{tab:eval-setup}
  \centering
  \begin{subtable}{0.96\linewidth}
  \caption{Testing methods.}
  \label{tab:eval-methods}
  \centering
  \small
  \setlength{\tabcolsep}{9pt}
  \ifdefined\TOSEMVERSION
  \renewcommand{\arraystretch}{1.05}
  \begin{tabular}{@{}p{0.29\linewidth}p{\dimexpr0.71\linewidth-2\tabcolsep\relax}@{}}
    \toprule
    Family & Methods (year) \\
    \midrule
    Graph-derived
      & \raggedright \mbox{Gamera (2023)}, \mbox{GSlicer (2025)} \tabularnewline
    \addlinespace[3pt]
    \mbox{Statement metamorphic}
      & \raggedright \mbox{GDBMeter (2023)}, \mbox{Gamera (2023)}, \mbox{GraphGenie (2024)},
        \mbox{GraspDB (2024)}, \mbox{GRev (2024)}, \mbox{Dinkel (2026)} \tabularnewline
    \addlinespace[3pt]
    Differential & GDsmith (2023) \\
    \bottomrule
  \end{tabular}
  \else
  \setlength{\tabcolsep}{4pt}
  \begin{tabular}{p{1.55cm}p{1.5cm}p{0.55cm}}
    \toprule
    Family & Source & Year \\
    \midrule
    \multirow{2}{=}{Graph-derived}
      & Gamera   & 2023 \\
      & GSlicer  & 2025 \\
    \addlinespace[3pt]
    \multirow{6}{=}{Statement metamorphic}
      & GDBMeter   & 2023 \\
      & Gamera     & 2023 \\
      & GraphGenie & 2024 \\
      & GraspDB    & 2024 \\
      & GRev       & 2024 \\
      & Dinkel     & 2026 \\
    \addlinespace[3pt]
    Differential & GDsmith & 2023 \\
    \bottomrule
  \end{tabular}
  \fi
\end{subtable}

  \par\medskip
  \begin{subtable}{0.74\linewidth}
  \caption{Database systems.}
  \label{tab:eval-systems}
  \centering
  \small
  \renewcommand{\arraystretch}{1.05}
  \setlength{\tabcolsep}{9pt}
  \begin{tabular}{@{}lcrr@{}}
    \toprule
    System & Version & Rank & Stars \\
    \midrule
    Neo4j & 2026.03.1 & 1 & 16.8k \\
    Memgraph & 3.9.0 & 13 & 4.2k \\
    FalkorDB & 4.16.0 & 22 & 4.7k \\
    \bottomrule
  \end{tabular}
\end{subtable}

\end{table}

Our study is organized around the following research questions:

\begin{itemize}[leftmargin=1.5em]
  \item \textbf{RQ1:} What does the raw output of the integrated methods look like, and how much of it corresponds to real defects?
  \item \textbf{RQ2:} What root causes underlie the false positives that dominate this raw output?
  \item \textbf{RQ3:} Which query patterns cause metamorphic follow-ups to collapse to the same execution plan, and which resist collapse?
\end{itemize}

\subsection{RQ1: Raw Reports Produced by the Integrated Methods}
\label{sec:study-rq1}

\paragraph*{Finding 1: reports are dominated by false positives}
After the campaigns finish, we observe that thousands of reports are produced,
which far exceeds the number of genuine defects a mature database is likely to contain.
We then inspect all the reports manually to distinguish false positives from reports of real bugs.
\reftable{rq1-report-funnel} shows the result of our inspection.

\begin{table}[!htbp]
  \caption{Report funnel by method and database. \emph{Raw} denotes reports
  before triage, \emph{FP} false positives, and \emph{Bugs} unique confirmed
  bugs. GDsmith runs one joint differential campaign across all three systems.}
  \label{tab:rq1-report-funnel}
  \centering
  \ifdefined\TOSEMVERSION\footnotesize\else\scriptsize\fi
  \renewcommand{\arraystretch}{1.04}

  \ifdefined\TOSEMVERSION
  {\sffamily\bfseries\footnotesize (a) Per-database campaigns}\par
  \vspace{0.3em}
  \fontsize{7.2}{8.5}\selectfont
  \setlength{\tabcolsep}{2.2pt}
  \renewcommand{\arraystretch}{1.08}
  \begin{tabular*}{\linewidth}{@{\extracolsep{\fill}}lrrrrrrrrr@{}}
    \toprule
    & \multicolumn{3}{c}{\textbf{Neo4j}}
    & \multicolumn{3}{c}{\textbf{Memgraph}}
    & \multicolumn{3}{c@{}}{\textbf{FalkorDB}} \\
    \cmidrule(lr){2-4}\cmidrule(lr){5-7}\cmidrule(l){8-10}
    \textbf{Method}
    & \textbf{Raw} & \textbf{FP} & \textbf{Bugs}
    & \textbf{Raw} & \textbf{FP} & \textbf{Bugs}
    & \textbf{Raw} & \textbf{FP} & \textbf{Bugs} \\
    \midrule
    Gamera     & 98  & 54  & \textbf{1} & 234 & 123 & \textbf{1} & 178 & 156 & \textbf{2} \\
    GDBMeter   & 0   & 0   & 0          & 12  & 0   & \textbf{1} & 134 & 134 & 0 \\
    GraphGenie & 54  & 54  & 0          & 133 & 133 & 0          & 527 & 249 & \textbf{1} \\
    GraspDB    & 40  & 40  & 0          & 0   & 0   & 0          & 35  & 35  & 0 \\
    GRev       & 9   & 9   & 0          & 60  & 13  & \textbf{1} & 10  & 2   & \textbf{1} \\
    GSlicer    & 128 & 102 & \textbf{1} & 362 & 203 & \textbf{1} & 412 & 342 & \textbf{1} \\
    Dinkel     & 22  & 0   & \textbf{3} & 36  & 12  & \textbf{3} & 46  & 36  & \textbf{1} \\
    \bottomrule
  \end{tabular*}

  \vspace{0.65em}
  \begin{minipage}{0.72\linewidth}
    \centering
    {\sffamily\bfseries\footnotesize (b) Joint differential campaign}\par
    \vspace{0.25em}
    \setlength{\tabcolsep}{4pt}
    \begin{tabular*}{\linewidth}{@{\extracolsep{\fill}}lrrr@{}}
      \toprule
      \textbf{Compared systems} & \textbf{Raw} & \textbf{FP} & \textbf{Bugs} \\
      \midrule
      Neo4j + Memgraph + FalkorDB & 1,584 & 1,439 & \textbf{2} \\
      \bottomrule
    \end{tabular*}
  \end{minipage}
  \else
  \begin{subtable}{\columnwidth}
    \centering
    \caption{Per-database report funnels.}
    \label{tab:rq1-report-funnel-per-db}
    \setlength{\tabcolsep}{3pt}
    \ifdefined\TOSEMVERSION
    \begin{tabular*}{\linewidth}{@{\extracolsep{\fill}}llccc@{}}
    \else
    \begin{tabular}{llccc}
    \fi
      \toprule
      \textbf{Method} & \textbf{DB} & \textbf{Rep.} & \textbf{FP} & \textbf{Unique Bug} \\
      \midrule
      Gamera     & Neo4j    & 98  & 54  & 1 \\
                 & Memgraph & 234 & 123 & 1 \\
                 & FalkorDB & 178 & 156 & 2 \\
      \addlinespace[1pt]
      GDBMeter   & Neo4j    & 0   & 0   & 0 \\
                 & Memgraph & 12  & 0   & 1 \\
                 & FalkorDB & 134 & 134 & 0 \\
      \addlinespace[1pt]
      GraphGenie & Neo4j    & 54  & 54  & 0 \\
                 & Memgraph & 133 & 133 & 0 \\
                 & FalkorDB & 527 & 249 & 1 \\
      \addlinespace[1pt]
      GraspDB    & Neo4j    & 40  & 40  & 0 \\
                 & Memgraph & 0   & 0   & 0 \\
                 & FalkorDB & 35  & 35  & 0 \\
      \addlinespace[1pt]
      GRev       & Neo4j    & 9   & 9   & 0 \\
                 & Memgraph & 60  & 13  & 1 \\
                 & FalkorDB & 10  & 2   & 1 \\
      \addlinespace[1pt]
      GSlicer    & Neo4j    & 128 & 102 & 1 \\
                 & Memgraph & 362 & 203 & 1 \\
                 & FalkorDB & 412 & 342 & 1 \\
      \addlinespace[1pt]
      Dinkel     & Neo4j    & 22  & 0   & 3 \\
                 & Memgraph & 36  & 12  & 3 \\
                 & FalkorDB & 46  & 36  & 1 \\
      \bottomrule
    \ifdefined\TOSEMVERSION\end{tabular*}\else\end{tabular}\fi
  \end{subtable}

  \vspace{0.25em}

  \begin{subtable}{\columnwidth}
    \centering
    \caption{GDsmith tri-database campaign.}
    \label{tab:rq1-report-funnel-gdsmith}
    \setlength{\tabcolsep}{4pt}
    \ifdefined\TOSEMVERSION
    \begin{tabular*}{\linewidth}{@{\extracolsep{\fill}}lccc@{}}
    \else
    \begin{tabular}{lccc}
    \fi
      \toprule
      \textbf{Compared Systems} & \textbf{Rep.} & \textbf{FP} & \textbf{Unique Bug} \\
      \midrule
      Neo4j / Memgraph / FalkorDB & 1,584 & 1,439 & 2 \\
      \bottomrule
    \ifdefined\TOSEMVERSION\end{tabular*}\else\end{tabular}\fi
  \end{subtable}
  \fi
\end{table}

For example,
\textsc{GDsmith} produces 1,439 false positives out of 1,584 raw reports in its tri-database differential campaign,
and \textsc{GraspDB} produces 40 of 40 on Neo4j and 35 of 35 on FalkorDB, making every report a false positive.
Overall, 3,136 of the 4,114 raw reports (76.2\%) are false positives, so most reports produced by the integrated methods do not correspond to real bugs.

\paragraph*{Finding 2: Reports of real bugs are dominated by duplicates}
Once false positives are set aside, 
the remaining reports of real bugs still number in the hundreds.
Of \textsc{GraphGenie}'s 527 raw reports on FalkorDB, 249 are false positives; the remaining 278 reports all refer to a single unique bug.
Similarly, \textsc{Gamera}'s 234 reports on Memgraph include 123 false positives, and the remaining 111 reports all refer to one bug.
Most reports of real bugs are therefore duplicates, and false-positive filtering must be followed by deduplication to determine the number of unique bugs.
Separating the few real defects from this flood of false positives and duplicates demands substantial manual effort,
which is often impractical.

\subsection{RQ2: Characterizing False Positives}
\label{sec:study-rq2}

To explain the false positives observed in RQ1, we inspect their causes and group them into four categories.
\emph{Impl.} denotes implementation-induced false positives, where the reference implementation does not correctly implement the method described in the paper.
\emph{Graph Sem.} denotes false positives related to the semantic constraints of graph pattern matching.
\emph{Null/Pipe} denotes false positives from incorrect assumptions about null values or clause pipelines.
\emph{Dialect} denotes false positives caused by dialect incompatibilities.
\reftable{fp-category-distribution} reports the distribution of these false positives across the eight methods.

\begin{table}[!htbp]
\caption{Distribution of false-positive reports by root-cause category.}
\label{tab:fp-category-distribution}
\centering
\renewcommand{\arraystretch}{1.04}
\ifdefined\TOSEMVERSION
\footnotesize
\setlength{\tabcolsep}{5.5pt}
\begin{tabular}{@{}lccccc@{}}
\else
\setlength{\tabcolsep}{6pt}
\resizebox{\columnwidth}{!}{%
\begin{tabular}{lccccc}
\fi
\toprule
\textbf{Method} &
\textbf{Impl. (\%)} &
\textbf{Graph Sem. (\%)} &
\textbf{Null/Pipe (\%)} &
\textbf{Dialect (\%)} &
\textbf{FP Reports} \\
\midrule
Gamera     & 10.8 & 75.6 & 0 & 13.6 & 333 \\
GDBMeter   & 35.8 & 0 & 0 & 64.2 & 134 \\
GraphGenie & 59.2 & 0 & 24.5 & 16.3 & 436 \\
GraspDB    & 32.0 & 18.7 & 34.7 & 14.6 & 75 \\
GRev       & 29.2 & 0 & 0 & 70.8 & 24 \\
GSlicer    & 11.9 & 43.0 & 41.1 & 4.0 & 647 \\
Dinkel     & 0 & 56.2 & 0 & 43.8 & 48 \\
GDsmith    & 4.9 & 51.1 & 31.0& 13.0 & 1,439 \\
\bottomrule
\end{tabular}%
\ifdefined\TOSEMVERSION\else
}
\fi
\end{table}

\paragraph*{Finding 3: Overlooked graph-semantic constraints cause false positives}
In \reftable{fp-category-distribution}, \emph{Graph Sem.} denotes false positives caused by test oracles that overlook the target database's graph-semantic constraints.
As described in \refsec{bg-semantic-constraints}, these constraints determine which mappings of a graph pattern are valid and how many result rows each mapping produces.
We find this cause in five of the eight methods.
It accounts for more than half of the false positives from \textsc{Gamera} (75.6\%), \textsc{Dinkel} (56.2\%), and \textsc{GDsmith} (51.1\%).
The oracles of these test methods implicitly assume particular semantic constraints when checking the query results of test cases.
If these assumptions conflict with the target system's actual choices, the oracles produce false positives.
To understand these false positives, we next examine how graph-semantic constraints differ across database systems.

\paragraph*{Finding 4: Graph-semantic constraints differ across database systems}
We examine how eight graph database systems that support Cypher implement the three constraints defined in \refsec{bg-semantic-constraints}.
Their choices form five combinations of constraint settings (\reftable{semantic-constraints}).

\begin{table}[H]
  \caption{Graph-semantic constraints in eight graph database systems.}
  \label{tab:semantic-constraints}
  \centering
  \footnotesize
  \ifdefined\TOSEMVERSION
    \setlength{\tabcolsep}{6pt}
  \else
    \setlength{\tabcolsep}{4pt}
  \fi
  \begin{tabular}{@{}llll@{}}
    \toprule
    System & Legality & Multiplicity & Reset boundary \\
    \midrule
    Neo4j       & no reuse   & single           & MATCH clause \\
    Memgraph    & no reuse   & length-amplified & MATCH clause \\
    ArcadeDB    & no reuse   & length-amplified & MATCH clause \\
    Apache AGE  & no reuse   & single           & comma-pattern \\
    NebulaGraph & no reuse   & single           & comma-pattern \\
    FalkorDB    & edge reuse & single           & n/a \\
    NeuG        & edge reuse & per-direction    & n/a \\
    Kuzu        & edge reuse & per-direction    & n/a \\
    \bottomrule
  \end{tabular}
\end{table}

First, legality differs: Neo4j, Memgraph, ArcadeDB~\cite{ArcadeDB}, Apache~AGE~\cite{ApacheAGE}, and NebulaGraph~\cite{wu2022nebula} reject same-edge reuse within the relevant scope, whereas FalkorDB, NeuG, and Kuzu~\cite{feng2023kuzu} permit it.
Second, multiplicity differs even when legality is fixed.
On a one-hop undirected self-loop, six systems emit a single witness, but NeuG and Kuzu emit two, one per traversal direction.
Under variable-length traversal, Memgraph and ArcadeDB amplify a single legal sequence into multiple witnesses, whereas the other no-reuse systems do not.
FalkorDB permits edge reuse yet emits single witnesses, confirming that these two constraints, legality and multiplicity, are separable.
Third, the reset boundary differs: Neo4j, Memgraph, and ArcadeDB refresh the uniqueness scope only at MATCH-clause boundaries, while Apache~AGE and NebulaGraph refresh it at comma-pattern boundaries; for engines that permit edge reuse the boundary is inapplicable.
The same query can therefore yield different row counts because systems differ in which relationship bindings they admit and how often they count each binding.

\begin{figure}[!htbp]
  \centering
  \ifdefined\TOSEMVERSION
  \includegraphics[page=1,width=\linewidth]{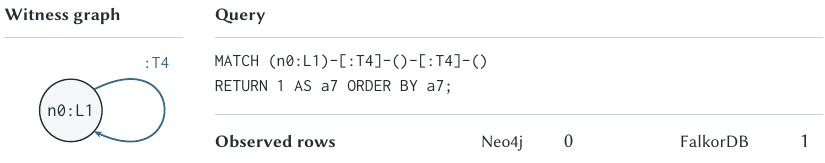}
  \else
  \begin{minipage}[c]{0.34\columnwidth}
    \centering
    \includegraphics[height=20mm]{figures/rq2-edge-uniqueness.pdf}\\[0.2em]
    {\scriptsize Witness graph}
  \end{minipage}
  \hfill
  \begin{minipage}[c]{0.55\columnwidth}
    \centering
    \small
    \begin{tabular}{lc}
    \toprule
    \textbf{DBMS} & \textbf{Rows} \\
    \midrule
    Neo4j & 0 \\
    FalkorDB & 1 \\
    \bottomrule
    \end{tabular}
  \end{minipage}

  \vspace{0.4em}
  \small
  \begin{tabular}{@{}l@{}}
  \toprule
  \textbf{GDsmith-style query} \\
  \midrule
  \texttt{MATCH (n0:L1)-[:T4]-()-[:T4]-()} \\
  \texttt{RETURN 1 AS a7 ORDER BY a7;} \\
  \bottomrule
  \end{tabular}
  \fi
  \caption{A differential false positive caused by relationship-uniqueness semantics.}
  \Description{A one-node graph with a self-loop labeled T4, a two-hop anonymous Cypher query over label L1 and relationship type T4, and a result table showing Neo4j with zero rows and FalkorDB with one row.}
  \label{fig:rq2-edge-uniqueness}
\end{figure}

\paragraph*{Case 1: relationship binding decides \emph{which} edges may fill a pattern}
A representative example from \textsc{GDsmith} is shown in \reffig{rq2-edge-uniqueness}.
The witness graph contains one node and one self-loop. 
\textsc{GDsmith} can generate an anonymous two-hop pattern and return only a constant value. 
On FalkorDB, the single self-loop can be reused for both relationship slots, so the query produces one row. 
On Neo4j, the same match is rejected because the engine enforces relationship uniqueness within the graph pattern, so the query returns no rows. 
Both edge-uniqueness semantics are valid, so the row-count mismatch is a false positive of the differential oracle rather than a defect in either system.

\FloatBarrier

\paragraph*{Case 2: enumeration multiplicity decides \emph{how many times} a match is reported}
A representative false positive from the metamorphic method \textsc{Dinkel} is shown in \reffig{rq2-path-multiplicity}.
The witness graph contains one node and two self-loops.
\textsc{Dinkel} rewrites a two-hop undirected expansion into two adjacent one-hop expansions and expects the source and follow-up to agree.
On Neo4j, both queries return two rows.
On Memgraph, the source returns four rows and the split returns eight, so the metamorphic oracle flags a mismatch.

\begin{figure}[!htbp]
  \centering
  \ifdefined\TOSEMVERSION
  \includegraphics[page=2,width=\linewidth]{figures/semantic-cases-tosem.pdf}
  \else
  \begin{minipage}[c]{0.34\columnwidth}
    \centering
    \includegraphics[height=24mm]{figures/rq2-path-multiplicity.pdf}\\[0.2em]
    {\scriptsize Witness graph}
  \end{minipage}
  \hfill
  \begin{minipage}[c]{0.55\columnwidth}
    \centering
    \small
    \begin{tabular}{lcc}
    \toprule
    \textbf{DBMS} & \textbf{Source} & \textbf{Split} \\
    \midrule
    Neo4j & 2 & 2 \\
    Memgraph & 4 & 8 \\
    \bottomrule
    \end{tabular}
  \end{minipage}

  \vspace{0.4em}
  \scriptsize
  \begin{tabular}{@{}ll@{}}
  \toprule
  \textbf{Source} & \texttt{MATCH p=(a:N)-[*2..2]-(a:N) RETURN relationships(p)} \\
  \textbf{Split} & \texttt{MATCH p=(a:N)-[*1..1]-()-[*1..1]-(a:N) RETURN relationships(p)} \\
  \bottomrule
  \end{tabular}
  \fi
  \caption{A metamorphic false positive caused by path-enumeration multiplicity.}
  \Description{A one-node graph with two self-loops labeled r1 and r2, two path queries representing a two-hop variable-length expansion and its split form, and a result table showing Neo4j with two rows for both queries and Memgraph with four rows for the source query and eight rows for the split query.}
  \label{fig:rq2-path-multiplicity}
\end{figure}

This mismatch is not simply a relationship-uniqueness violation.
An explicit two-hop pattern with named relationships does not admit the binding $r=s$ on either Neo4j or Memgraph in this graph, so the legal binding set is unchanged.
The discrepancy instead comes from bag-semantic path enumeration across a variable-length segment boundary.
Neo4j treats an undirected self-loop as a single traversal direction, whereas Memgraph enumerates both directions independently within each one-hop segment, multiplying the final row count.
The rewrite therefore preserves \emph{which} paths are legal but changes \emph{how many times} each legal path is emitted.
Because both engines apply their respective enumeration rules consistently, the report is a false positive of the metamorphic oracle rather than a defect in either system.

\paragraph*{Finding 5: Inconsistent application of graph-semantic constraints reveals real bugs}
When inspecting reports, we also discover real bugs caused by these semantic constraints.
\reffig{rq2-binding-contradiction} shows a real FalkorDB bug we discovered.
The witness graph contains two nodes, two reciprocal edges, and one self-loop.
From node 1 back to node 1, the exact-length form, the explicit two-hop form, and the split form should describe the same matching space.
Neo4j and Memgraph return two rows on all three forms.
FalkorDB also returns two rows on the exact-length form, but five on the explicit and split forms.

\begin{figure}[!htbp]
  \centering
  \ifdefined\TOSEMVERSION
  \includegraphics[page=3,width=\linewidth]{figures/semantic-cases-tosem.pdf}
  \else
  \begin{minipage}[c]{0.34\columnwidth}
    \centering
    \resizebox{0.82\linewidth}{!}{\input{figures/rq2-binding-contradiction-witness}}\\[0.2em]
    {\scriptsize Witness graph}
  \end{minipage}
  \hfill
  \begin{minipage}[c]{0.55\columnwidth}
    \centering
    \scriptsize
    \setlength{\tabcolsep}{4.5pt}
    \begin{tabular}{lccc}
    \toprule
    \textbf{DBMS} & \textbf{Exact} & \textbf{Explicit} & \textbf{Split} \\
    \midrule
    Neo4j & 2 & 2 & 2 \\
    Memgraph & 2 & 2 & 2 \\
    FalkorDB & 2 & 5 & 5 \\
    \bottomrule
    \end{tabular}
  \end{minipage}

  \vspace{0.4em}
  \scriptsize
  \begin{tabular}{@{}lp{0.76\columnwidth}@{}}
  \toprule
  \textbf{Exact} &
  {\ttfamily MATCH p=(a:N \{id:1\}) -[*2..2]- (b:N \{id:1\}) \newline
  RETURN [rel IN relationships(p) | rel.eid] AS eids} \\
  \textbf{Explicit} &
  {\ttfamily MATCH (a:N \{id:1\}) -[r]- (m) -[s]- (b:N \{id:1\}) \newline
  RETURN r.eid AS reid, s.eid AS seid} \\
  \textbf{Split} &
  {\ttfamily MATCH p=(a:N \{id:1\}) -[*1..1]- () -[*1..1]- (b:N \{id:1\}) \newline
  RETURN [rel IN relationships(p) | rel.eid] AS eids} \\
  \bottomrule
  \end{tabular}
  \fi
  \caption{A real bug caused by contradictory relationship-uniqueness enforcement.}
  \Description{A two-node witness graph with reciprocal edges r1 and r2 and a self-loop r3 on node 1, together with a result table showing Neo4j and Memgraph returning two rows on the exact, explicit, and split two-hop forms, while FalkorDB returns two rows on the exact form and five rows on the explicit and split forms.}
  \label{fig:rq2-binding-contradiction}
\end{figure}

Unlike Case~2, this discrepancy changes the legal binding set itself rather than only the multiplicity.
Repeated-edge bindings such as \texttt{[r1,r1]} and the self-loop self-pair \texttt{[r3,r3]} are excluded by the exact-length form but admitted by the explicit and split forms on FalkorDB.
This is a real bug because the system contradicts itself about whether the same edge may occupy two relationship slots in one match.

Taken together, RQ2 shows that graph-semantic constraints are a common source of false positives across multiple testing methods.
When different systems make different but internally self-consistent choices, the report is a false positive;
when one system gives contradictory answers to the same constraint across equivalent forms, the mismatch is a real defect.

\subsection{RQ3: Characterizing Plan Collapse}
\label{sec:study-rq3}

Using the execution plans recorded by \sys, we study how often equivalent query pairs compile to the same plan, why this happens, and which transformations retain different plans.

\paragraph*{Finding 6: 5--34\% of equivalent query pairs lose their ability to detect bugs}
\reftable{plan-status} reports the same-plan ratio for comparable source/follow-up pairs. 
For each method-database combination we repeat the campaign under three independent seeds, each run stopping once it collects 5{,}000 comparable pairs whose two queries both execute and expose a plan. Each cell reports the mean fraction sharing a plan hash across the three runs. The three runs agree within 1.3 percentage points.
As the table shows, pairs that compile to the same plan appear in every equivalence-based metamorphic method, 
ranging from roughly 5\% to 34\% of comparable pairs.
Notably, GraspDB's same-plan ratio rises from 5.5\% on Neo4j to 33.2\% on FalkorDB, a gradient that reflects dialect fit rather than optimizer strength.
FalkorDB and Memgraph accept fewer of its write-bearing variants, so its comparable pairs skew toward collapse-prone forms.

\begin{figure}[!htbp]
  \centering
  \begin{minipage}{\linewidth}
    \captionsetup{type=table}
    \caption{Same-plan ratios by method and database, averaged over three runs.}
\label{tab:plan-status}
\centering
\renewcommand{\arraystretch}{1.04}
\setlength{\tabcolsep}{6pt}
\begin{tabular}{lccc}
\toprule
\textbf{Method} & \textbf{Neo4j} & \textbf{Memgraph} & \textbf{FalkorDB} \\
\midrule
\textsc{GraphGenie} & 11.29\% & 15.34\% & 18.94\% \\
\textsc{GRev}       & 13.71\% & 5.79\%  & 19.76\% \\
\textsc{GraspDB}    & 5.50\%  & 26.67\% & 33.24\% \\
\textsc{Dinkel}     & 27.44\% & 17.44\% & 34.21\%\\
\bottomrule
\end{tabular}

  \end{minipage}
  \par\medskip
  \ifdefined\TOSEMVERSION
  \includegraphics[width=0.90\linewidth]{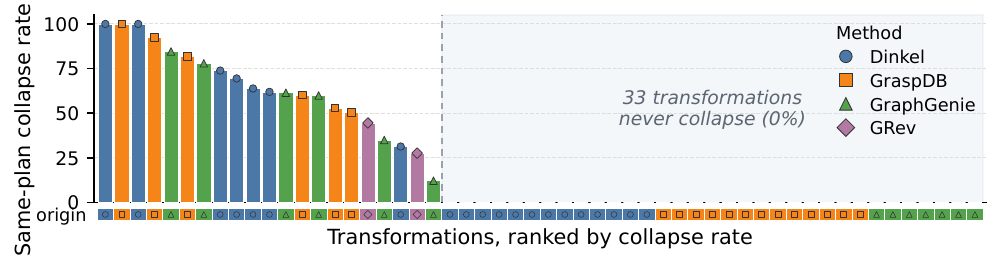}
  \else
  \includegraphics[width=\columnwidth]{figures/plan-collapse-per-transformation.pdf}
  \fi
  \caption{Same-plan collapse rates of 54 query transformations on Neo4j.}
  \Description{A ranked bar chart of 54 transformations by same-plan collapse rate, with each method represented by a distinct color and marker shape. A rug beneath the axis shows the originating method of every transformation, including the 33 zero-collapse ones contributed by Dinkel, GraspDB, and GraphGenie.}
  \label{fig:plan-collapse-per-transformation}
\end{figure}

The per-method ratios in \reftable{plan-status} average over transformations that individually behave very differently.
\reffig{plan-collapse-per-transformation} resolves the Neo4j campaign into its 54 equivalence-preserving transformations, ranked by collapse rate and encoded by the color and marker shape of their originating method; a collapse-resistant transformation leaves no bar, so a rug beneath the axis marks its method.

The transformations divide into those that collapse and the 33 of 54 that never do.
Which side a transformation falls on follows its shape rather than its source.
A single method can include transformations that collapse to the source plan and others that retain distinct plans.

\S\ref{sec:bg-query-optimization} distinguished the two optimization classes that can erase a follow-up's distinction: 
\emph{rule-based rewriting}, 
which canonicalizes a query independently of the data, 
and \emph{cost-based optimization}, 
which searches for the cheapest plan under cardinality statistics.
We present one collapse driven by each class and a third transformation that neither can collapse,
verified on Neo4j with its default planner over a graph with a skewed label distribution where $|A|{=}5$, $|B|{=}500$, and $|C|{=}50$, with $500$ \texttt{R}- and $500$ \texttt{S}-edges.

\paragraph*{Case A: rule-based rewriting collapses syntactic differences regardless of the data}
The simplest collapses are structural: the rewrite phase canonicalizes surface variation before any cost is considered.
In \reffig{rq2-rule-collapse},
the follow-up reverses the written traversal direction from \texttt{(a)-[:R]->(b)} to \texttt{(b)<-[:R]-(a)} and appends a tautological \texttt{WHERE true}, both of which are transformations that equivalence-based methods routinely apply.
Direction normalization and dead-predicate elimination are semantics-preserving,
so Neo4j folds the two queries to an identical operator tree, and both return $500$.
Driven by rewrite rules rather than statistics, this collapse is data-independent:
the queries share a plan on \emph{any} instance.

\begin{figure}[!htbp]
  \centering
  \scriptsize
  \begin{tabular}{@{}p{\columnwidth}@{}}
  \toprule
  \textbf{Source query} \\
  \texttt{MATCH (a:A)-[:R]->(b:B) RETURN count(*) AS cnt} \\
  \midrule
  \textbf{Follow-up: direction reversal + dead predicate} \\
  \texttt{MATCH (b:B)<-[:R]-(a:A) WHERE true} \\
  \texttt{RETURN count(*) AS cnt} \\
  \midrule
  \textbf{Shared Neo4j \texttt{EXPLAIN} operator tree} \\
  \texttt{ProduceResults | cnt} \\
  \texttt{\ \ EagerAggregation | count(*) AS cnt} \\
  \texttt{\ \ \ \ Filter | b:B} \\
  \texttt{\ \ \ \ \ \ Expand(All) | (a)-[:R]->(b)} \\
  \texttt{\ \ \ \ \ \ \ \ NodeByLabelScan | a:A} \\
  \bottomrule
  \end{tabular}
  \caption{A rule-based plan-collapse case.}
  \Description{A source single-hop pattern, a follow-up that reverses the relationship direction and adds a tautological WHERE true predicate, and the identical Neo4j operator tree shared by both queries.}
  \label{fig:rq2-rule-collapse}
\end{figure}

\paragraph*{Case B: cost-based optimization collapses join orders chosen from the data}
In \reffig{rq2-cost-collapse}, a two-hop chain is written as two comma-separated patterns, 
and the follow-up lists them in the opposite order.

\begin{figure}[!htbp]
  \centering
  \scriptsize
  \begin{tabular}{@{}p{\columnwidth}@{}}
  \toprule
  \textbf{Source query} \\
  \texttt{MATCH (a:A)-[:R]->(b:B), (b)-[:S]->(c:C)} \\
  \texttt{RETURN count(*) AS cnt} \\
  \midrule
  \textbf{Follow-up: comma-pattern reordering} \\
  \texttt{MATCH (b:B)-[:S]->(c:C), (a:A)-[:R]->(b)} \\
  \texttt{RETURN count(*) AS cnt} \\
  \midrule
  \textbf{Shared Neo4j \texttt{EXPLAIN} operator tree} \\
  \texttt{ProduceResults | cnt} \\
  \texttt{\ \ EagerAggregation | count(*) AS cnt} \\
  \texttt{\ \ \ \ Filter | c:C} \\
  \texttt{\ \ \ \ \ \ Expand(All) | (b)-[:S]->(c)} \\
  \texttt{\ \ \ \ \ \ \ \ Filter | b:B} \\
  \texttt{\ \ \ \ \ \ \ \ \ \ Expand(All) | (a)-[:R]->(b)} \\
  \texttt{\ \ \ \ \ \ \ \ \ \ \ \ NodeByLabelScan | a:A} \\
  \bottomrule
  \end{tabular}
  \caption{A cost-based plan-collapse case.}
  \Description{A source two-hop chain written as two comma-separated patterns, a follow-up listing the same patterns in reverse order, and the shared Neo4j operator tree anchored at label A expanding to B then C.}
  \label{fig:rq2-cost-collapse}
\end{figure}

Both compile to the same plan,
which is anchored by a \texttt{NodeByLabelScan} on \texttt{A} and expands \texttt{A}$\to$\texttt{B}$\to$\texttt{C}, and both return $500$.
The anchor is set by statistics,
not syntax: \texttt{A} is the rarest label,
so the planner starts there and drives expansion from the smallest endpoint regardless of how the patterns were written.
A control confirms the data dependence:
on an isomorphic chain \texttt{P}$\to$\texttt{Q}$\to$\texttt{Z} whose rare label sits at the other end, where $|Z|{=}5$,
the same reordering instead collapses to a plan anchored at \texttt{Z}.
The convergence point moves with the cardinality estimates, which is the signature of cost-based rather than rule-based collapse.

Both rule-based rewriting and cost-based optimization can therefore eliminate the plan differences introduced by equivalence-preserving transformations.
Because equivalence-based metamorphic methods detect bugs via inconsistent results, executing both queries against the same graph state with the same plan makes their results identical by construction.
Such pairs consume testing resources without revealing defects; if their shared plan contains a bug, both queries can return the same incorrect result and the oracle remains silent.

\paragraph*{Finding 7: Some query transformations resist plan collapse}
\textsc{GraspDB} builds follow-ups by injecting \emph{net-zero write sequences}: sequences of write operations that leave the graph state unchanged after completion, such as creating and then deleting a node or assigning a property on a \texttt{null} target.
These sequences preserve the intended query results and final graph state while adding write operators to the follow-up plan.
In \reffig{rq2-write-resist}, the follow-up extends a write source with a \texttt{CREATE c0 \dots DELETE c0} no-op.
The added sequence creates a temporary node and deletes it before the query completes, without modifying existing graph data; the final graph state is therefore the same as after the source query.
Neo4j nevertheless retains the \texttt{Create} and \texttt{Delete} operators and inserts an \texttt{Eager} barrier to enforce read/write ordering, leaving the follow-up plan structurally distinct from the source.
Rule-based rewriting cannot drop a write operator, as write semantics forbid treating it as dead code, and cost-based optimization cannot reorder it past the read it conflicts with.

\begin{figure}[!htbp]
  \centering
  \scriptsize
  \begin{tabular}{@{}p{\columnwidth}@{}}
  \toprule
  \textbf{Source query} \\
  \texttt{MATCH (a:A) SET a.k = a.k + 1} \\
  \texttt{RETURN count(*) AS cnt} \\
  \midrule
  \textbf{Follow-up: injected net-zero write} \\
  \texttt{MATCH (a:A) SET a.k = a.k + 1} \\
  \texttt{CREATE (c0 \{id:-1\}) DELETE c0} \\
  \texttt{RETURN count(*) AS cnt} \\
  \midrule
  \textbf{Source plan} \\
  \texttt{ProduceResults | cnt} \\
  \texttt{\ \ EagerAggregation | count(*) AS cnt} \\
  \texttt{\ \ \ \ SetProperty | a.k = a.k + 1} \\
  \texttt{\ \ \ \ \ \ NodeByLabelScan | a:A} \\
  \midrule
  \textbf{Follow-up plan: write operators survive} \\
  \texttt{ProduceResults | cnt} \\
  \texttt{\ \ EagerAggregation | count(*) AS cnt} \\
  \texttt{\ \ \ \ Delete | c0} \\
  \texttt{\ \ \ \ \ \ Eager | read/delete conflict} \\
  \texttt{\ \ \ \ \ \ \ \ Create | (c0 \{id:-1\})} \\
  \texttt{\ \ \ \ \ \ \ \ \ \ SetProperty | a.k = a.k + 1} \\
  \texttt{\ \ \ \ \ \ \ \ \ \ \ \ NodeByLabelScan | a:A} \\
  \bottomrule
  \end{tabular}
  \caption{A collapse-resistant case.}
  \Description{A source write query setting a property, a follow-up adding a create-then-delete no-op write, and two distinct Neo4j operator trees showing that the follow-up retains Create, Delete, and Eager operators.}
  \label{fig:rq2-write-resist}
\end{figure}

This is why write-oriented metamorphic testing resists the plan collapse that read-only equivalence rewrites suffer, and it underlies the method complementarity exploited in the bug-discovery evaluation in \S\ref{sec:eval-bugdiscovery}.

\subsection{Guidelines from the Study}
\label{sec:motivation}
We distill the study into the following guidelines for both integrated testing and future test-method design.

\paragraph*{G1: Preserve process-level data for postprocessing}
Our analysis depends on information that raw oracle verdicts do not contain, including construction traces, oracle assumptions, execution plans, and observed outputs.
An integrated platform should therefore preserve these process-level artifacts as first-class report fields, so postprocessing can classify false positives, distinguish semantic ambiguity from bugs, and deduplicate reports by root cause.

\paragraph*{G2: Separate semantic false positives from real bugs}
In semantic regions where systems may validly differ, cross-system disagreement alone is weak bug evidence.
A stronger signal is intra-system contradiction: the same system should not disagree across equivalent forms that exercise the same semantic dimension.

\paragraph*{G3: Treat graph semantics as oracle preconditions}
Graph-pattern oracles encode semantic choices, including legality, witness multiplicity, and reset boundaries.
A method should declare these assumptions explicitly and check them against the target semantic profile before treating a mismatch as a failure.

\paragraph*{G4: Triage optimizer-collapsed follow-ups}
For equivalence-based metamorphic testing, semantic preservation is not enough: a source/follow-up pair that compiles to the same plan on the same graph state has no oracle power.
The execution pipeline should therefore identify same-plan pairs and remove them from bug-oriented inspection.

\paragraph*{G5: Design transformations that survive optimization}
Future metamorphic methods should consider whether their follow-ups remain distinguishable after rule-based and cost-based optimization.
Collapse-resistant transformations, such as net-zero writes, provide complementary signal because they preserve semantics while still exercising different execution paths.

\section{Bug Postprocessing and Process-Level Analysis}
\label{sec:whitebox}

The study of \S\ref{sec:study} provides guidelines for integrated testing.
We follow the guidelines to mitigate the two blind spots described above.
The design mainly consists of two components: a unified postprocessing workflow and a plan-aware triage.

\subsection{Overview}
\label{sec:whitebox-overview}
\begin{figure}[!htbp]
\centering
\includegraphics[width=\textwidth]{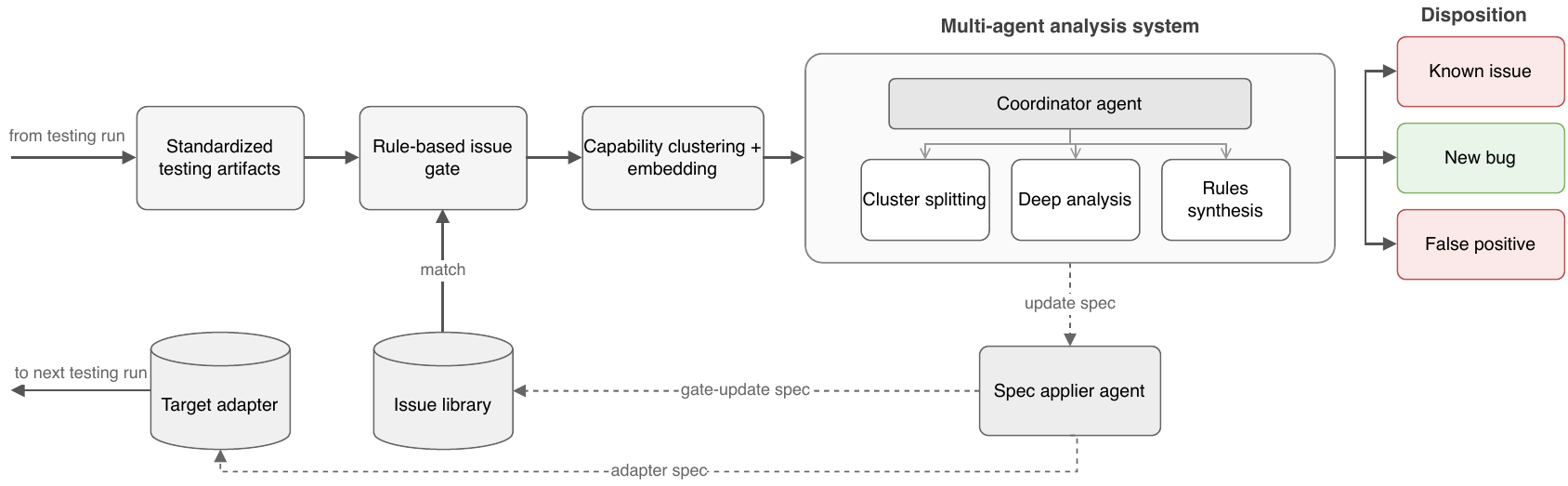}
\caption{\sys's postprocessing pipeline. Standardized artifacts pass through issue gating, capability-aware clustering, and multi-agent analysis; confirmed dispositions update the issue library or target adapter.}
\Description{Overview diagram with standardized testing artifacts flowing through a rule-based issue gate, capability-label clustering, and an LLM analysis pipeline. The analysis pipeline produces disposition decisions and sends feedback to both the issue library and the target adapter.}
\label{fig:postprocess-workflow}
\end{figure}

As discussed in \S\ref{sec:introduction}, running integrated methods for hours can produce thousands of bug reports,
including many false positives and duplicate reports.
This motivates a postprocessing pipeline that automates these tasks and keeps human effort manageable.
The pipeline consists of three stages: noise reduction, report analysis, and feedback mechanisms, as shown in \reffig{postprocess-workflow}.

The pipeline first takes standardized testing artifacts as input,
consisting of a raw report with an oracle violation record and relevant process-level data, such as execution plans, collected by the testing framework.
Reports then pass through three sequential stages. 
A rule-based issue gate matches incoming reports against a known-issue library and directly classifies any hits. 
Remaining reports are grouped by capability labels and then refined using report embeddings. 
The clustered reports then enter a multi-agent analysis system, 
in which a coordinator agent dispatches three sub-agents to perform in-depth analysis and produces,
for each cluster, a disposition decision: known issue, false positive, or new bug.
If the decision is a known issue, a gate-update specification is synthesized and registered back into the issue library.
If the decision is a dialect-related false positive, an adapter specification is synthesized and returned to the target adapter of \S\ref{sec:system}, where it is applied automatically in subsequent testing.
This section details the multi-agent analysis system and the feedback mechanisms.

\subsection{Multi-Agent Analysis System}
\label{sec:whitebox-multi-agent}

Relying on an LLM to make disposition decisions during postprocessing raises two challenges.
The first is \textbf{context capacity}: 
raw reports carry sizable construction traces, execution plans, and target outputs.
Performing full deep analysis on every report would exhaust the context window.   
The second is \textbf{auditability}:
an LLM may produce plausible disposition judgments, 
but without explicit evidence requirements it is difficult for analysts to determine whether a judgment is well supported.
\sys addresses these two challenges at separate layers of its design.
All agents run on Claude Code backed by the \texttt{claude-sonnet-4-6} snapshot under its default decoding, and the three sub-agents share this backend, differing only in their stage prompt.

\paragraph*{On-demand sub-agents}
Mirroring how human analysts work,
we design three sub-agents to perform the three sub-tasks of report analysis.
The \textbf{cluster-splitting agent} assesses the internal homogeneity of each cluster by comparing two sampled anchors, the medoid with the smallest average cosine distance to the rest and the outlier farthest from that medoid;
if the cluster is heterogeneous, the agent splits it into two sub-clusters for further analysis.
The \textbf{deep-analysis agent} is responsible for deep analysis and producing verifiable evidence, 
and it operates through two established debugging techniques. 
The first is delta debugging~\cite{zeller2002simplifying}, which minimizes the witness graph and query while preserving the failure signal,
stripping away structures unrelated to the failure;
to enable the LLM to perform delta debugging effectively and efficiently, 
we distill human debugging experience into a skill that specifies the reduction objectives, decision criteria, and iteration strategy. 
The second is dynamic debugging, where the agent attaches a source-level debugger to a debug build of the system under test to inspect its runtime state.
To classify a report as a bug, 
the agent must provide both dynamic-debugging results and reproducible runtime evidence that links the observed failure to a concrete implementation path for human confirmation.
Notably, we distill expert knowledge derived from the study of \S\ref{sec:study} into the skill that guides this agent to identify whether a report actually represents a semantic-ambiguity false positive.
The \textbf{rules-synthesis agent} is activated only after a disposition has been confirmed,
and produces a gate rule or a dialect adapter specification.

\paragraph*{Disposition-specific acceptance criteria}
To ensure that the sub-agents adhere to these criteria, a coordinator agent audits the output of each sub-agent and triggers a re-run on demand.
Only once the evidence criteria of a disposition are met does the agent system commit to a decision, which falls into one of three categories: \emph{known issue}, \emph{new bug}, or \emph{false positive}.
Classifying a report as a \emph{known issue} or \emph{new bug} requires the agent to submit two artifacts.
The first is evidence, consisting of dynamic-debugging results together with supporting documentation, that establishes the behavior as a genuine defect.
The second is a gating rule for the feedback mechanism that prevents such reports from raising a signal in future runs; each rule carries both positive and negative cases, specifying what should and should not be filtered.
A \emph{false positive} is further attributed to one of two causes.
A \emph{dialect false positive} requires a dialect adapter specification together with alignment evidence showing that the adaptation preserves the testing method as faithfully as possible.
An \emph{oracle false positive}, which arises from semantic ambiguity, requires evidence that the observed database behavior is valid, established with the analysis skill described above.

\subsection{Feedback Mechanisms}
\label{sec:whitebox-feedback}

\sys feeds back only confirmed disposition results, and routes them according to the postprocessing guidelines of \S\ref{sec:motivation}.
The study distinguishes two feedback paths: dialect false positives require changing how requests are expressed, whereas confirmed bug families and duplicate reports require suppressing repeated analysis of the same root cause.
A dialect false positive updates the target adapter, because later executions should express the same testing intent in the target's dialect or reject an unsupported request shape.
A confirmed bug family or duplicate report updates the issue gate, because the input is valid but its signal should no longer trigger deep analysis.
These two outcomes define two independent feedback channels:
\emph{request-level adapter updates} and \emph{root-cause-level gate updates}.

\paragraph*{Request-level adapter update}
We model dialect adaptation as a request-level transformation task.
As discussed in \S\ref{sec:system},
a request of a test case can be abstracted into three components:
the target database, the graph state to be constructed, and the query sequence.
Before a request is sent to the database,
the target adapter transforms this combined information into a target-specific request.
Given confirmed dialect evidence, the updated adapter may rewrite statements,
lower graph construction into a target-specific form,
mark expected errors, or reject unsupported request shapes.
The key point is that adaptation operates on the full request rather than on an isolated syntax fragment.
When a false positive arises from the target dialect,
the adapter can adjust the request expression while preserving the testing intent,
instead of simply disabling the relevant feature.

The multi-agent analysis system synthesizes candidate adaptation rules,
which are then applied by the adapter's deterministic rule pipeline.
Any adaptation must preserve the graph state and statement relation;
if semantic equivalence cannot be preserved,
\sys disables or modifies the request shape rather than silently executing a test with a changed intent.
Each update is applied only within the dialect boundary supported by the evidence,
and is not generalized to superficially similar query classes.

\paragraph*{Root-cause-level gate update}
Root-cause-level gate feedback serves a different purpose:
multiple reports can stem from the same root cause.
We realize this with an issue library, where reports that share the same root cause are grouped into one issue family.
In the issue gate, each family is associated with multiple rules,
so that different manifestations of the same root cause are recognized as one issue and then filtered out.

When a report reaches this channel,
the multi-agent analysis system first checks whether it maps to an existing issue family;
if not, it registers a new family.
Because stable and concise root-cause analysis is hard for an agent to produce,
we do not aim to use the root cause it submits for direct bug fixing, but only for noise reduction.
The rules therefore need only capture the characteristics that group reports into the same issue family.

The two channels are not interchangeable, and \sys keeps them distinct: misrouting a confirmed bug family or duplicate report to the adapter can alter a valid input and mask future bugs, while misrouting a dialect false positive to the gate leaves the frontend generating invalid requests.
Because both updates enter existing pipelines as rule entries,
supporting a new target or issue type requires extending the corresponding rule library, without modifying the generator, adapter, and filter together.
This shifts human effort from inspecting individual bugs to auditing rules grouped by family.

\subsection{Plan-Aware Triage of Metamorphic Pairs}
\label{sec:whitebox-triage}

\refsec{study-rq3} showed that equivalence-based metamorphic methods routinely produce source/follow-up pairs that compile to the same plan on the same graph state, and that such a pair cannot expose a defect: an identical plan makes the two results equal by construction.
These pairs waste execution budget on tests with no oracle power.
For each equivalence pair, \sys first obtains the execution plans using \texttt{EXPLAIN} and compares them.
Only pairs with different plans are executed to compare their query results.

This yields three dispositions.
A pair with the \emph{same plan} on the same graph state is redundant by construction and is dropped without executing either query, reclaiming the budget quantified in \refsec{study-rq3}.
A pair with \emph{different plans but the same result} passes the oracle, yet its divergent plans on identical inputs hint at a missed optimization; \sys records the pair's execution times so that persistent gaps can be surfaced as performance issues rather than silently discarded.
A pair with \emph{different plans and different results} keeps its full oracle power and is retained for inspection as a candidate logic bug.

Only the same-plan class removes work, and it weakens no oracle: a pair whose queries share a plan on the same graph state cannot disagree, so skipping it forfeits no discoverable bug.
Because the triage reads only the plan and result already on the shared execution path described in \refsec{system}, every integrated equivalence-based method benefits without method-specific changes.

\section{Evaluation}
\label{sec:evaluation}

This section evaluates \sys and guidelines from \S\ref{sec:study} along two directions: whether the unified postprocessing workflow reduces analytical workload and whether the platform discovers real defects in a previously untested system.

\subsection{Experimental Setup}
\label{sec:eval-setup}

\paragraph*{Integrated methods and target systems}
We instantiate \sys with the same eight methods listed in \reftable{eval-setup}(\subref*{tab:eval-methods}) as in the study of \S\ref{sec:study}.
Both RQs additionally target NeuG 0.1.1~\cite{NeuG}, which had 118 GitHub stars in our June 2026 snapshot and was not previously exercised by any testing method.
We likewise run it with its default query optimizer enabled.

\paragraph*{Execution environment and metrics}
All campaigns reuse the unified containerized deployment of \S\ref{sec:study-setup}: each $(\textit{method}, \textit{system})$ pair runs in an isolated container against a separate database instance under the same 12-hour budget and timeout policy, with seeds and process-level artifacts recorded for replay.
We report outcomes at the same three levels that \S\ref{sec:study-setup} defines: \emph{raw report}, \emph{false positive}, and \emph{unique bug}.

\paragraph*{Validation protocol}
\sys's post-testing workflow first produces, for each raw report, an initial analysis that includes a false-positive assessment and a candidate root-cause hypothesis, serving only as an aid for human reviewers.
Two authors then independently inspect each report and reach a verdict, resolving disagreements through discussion.
For reports judged to be real bugs, we check whether the issue has already been reported and submit unreported cases to the developers for confirmation.

\subsection{Research Questions}
\label{sec:eval-rq}

Our evaluation is organized around the following research questions:

\begin{itemize}[leftmargin=1.5em]
  \item \textbf{RQ4:} To what extent does \sys's unified postprocessing workflow reduce the analytical overhead of raw testing output, and how faithfully do its automated dispositions agree with human judgment?

  \item \textbf{RQ5:} Does the unified platform enable practical bug discovery in a graph database system that has not been systematically exercised by prior work?
\end{itemize}

\FloatBarrier
\subsection{RQ4: Workload Reduction via Unified Postprocessing}
\label{sec:eval-workload}

As illustrated in \S\ref{sec:whitebox-overview}, 
reports found in \sys are processed by the unified postprocessing workflow.
The workflow is organized in batches. Within each batch, raw reports first pass through an initial gating step, 
are clustered by root cause, examined through representative cases to yield deep-analysis units, and finally aggregated into bug families. 
We define workload reduction as the relative decrease from the raw report count to the number of items that ultimately require human inspection.
A workflow that discards reports blindly would maximize this reduction, so we evaluate it on two axes:
how much it compresses the raw output, and how faithfully its automated dispositions agree with the human verdicts of \S\ref{sec:eval-setup}.
\reftable{rq3-workload} reports three representative batches on NeuG, to illustrate the per-batch reduction behavior.

\begin{table}[H]
  \caption{Per-batch workload reduction achieved by \sys's unified postprocessing workflow on NeuG, shown for three representative batches of the continuous campaign.}
  \label{tab:rq3-workload}
  \centering
  \renewcommand{\arraystretch}{1.06}
  \ifdefined\TOSEMVERSION
  \footnotesize
  \setlength{\tabcolsep}{3.5pt}
  \begin{tabular}{@{}lrrrrrrrr@{}}
    \toprule
    & \multicolumn{1}{c}{\textbf{Input}}
    & \multicolumn{1}{c}{\textbf{Filter}}
    & \multicolumn{2}{c}{\textbf{Analysis}}
    & \multicolumn{3}{c}{\textbf{Outputs}}
    & \multicolumn{1}{c}{\textbf{Workload}} \\
    \cmidrule(lr){2-2}
    \cmidrule(lr){3-3}
    \cmidrule(lr){4-5}
    \cmidrule(lr){6-8}
    \cmidrule(lr){9-9}
    \textbf{Batch} & \textbf{Raw} & \textbf{Kept}
      & \textbf{Clusters} & \textbf{Units}
      & \textbf{Adapter} & \textbf{Gate} & \textbf{Bugs}
      & \textbf{Reduction} \\
    \midrule
    \#1 & 26 & 13 & 3 & 4 & 2 & 1 & 1 & 84.6\% \\
    \#2 & 17 & 13 & 3 & 3 & 1 & 2 & 0 & 82.4\% \\
    \#3 & 12 &  9 & 3 & 4 & 1 & 2 & 1 & 66.7\% \\
    \bottomrule
  \end{tabular}
  \else
  \setlength{\tabcolsep}{4pt}
  \resizebox{\columnwidth}{!}{%
  \begin{tabular}{lcccccccc}
    \toprule
    \textbf{Scope} & \textbf{Raw} & \textbf{After} & \textbf{Clusters} & \textbf{Deep} & \textbf{Adapter} & \textbf{Gate} & \textbf{Bug} & \textbf{Workload} \\
                   & \textbf{reports} & \textbf{gate} &       & \textbf{units} & \textbf{rules} & \textbf{specs} & \textbf{families} & \textbf{reduction} \\
    \midrule
    Batch \#1 & 26 & 13 & 3 & 4 & 2 & 1 & 1 & 84.6\% \\
    Batch \#2 & 17 & 13 & 3 & 3 & 1 & 2 & 0 & 82.4\% \\
    Batch \#3 & 12 &  9 & 3 & 4 & 1 & 2 & 1 & 66.7\% \\
    \bottomrule
  \end{tabular}%
  }
  \fi
\end{table}

\paragraph*{The postprocessing workflow reduces analytical workload by 66.7\%--84.6\% across three batches}
Batch~\#1 gates 26 raw reports down to 13 and clusters them into 3 root-cause clusters,
which expand into 4 deep-analysis units mapped to 2 dialect-adapter rules, 1 gate spec, and 1 bug family, an 84.6\% reduction;
Batch~\#2 gates 17 raw reports down to 13 and clusters them into 3 clusters that map one-to-one to 3 units,
yielding 1 adapter rule and 2 gate specs, an 82.4\% reduction;
Batch~\#3 gates 12 raw reports down to 9 and clusters them into 3 clusters that expand into 4 units,
yielding 1 adapter rule, 2 gate specs, and 1 bug family, a 66.7\% reduction, as shown in \reftable{rq3-workload}.
Each cluster is examined through its centroid and an outlier representative:
a homogeneous cluster distills to a single unit, whereas a cluster that mixes two root causes splits into two,
so the number of deep-analysis units lies between the cluster count and twice that count,
and the adapter, gate, and bug columns sum to the unit count of each batch.
In aggregate, the workflow compresses the volume of items requiring human inspection to roughly one sixth to one third of the original raw report count, directly addressing the redundancy noise quantified in the study of \S\ref{sec:study}.

\paragraph*{The multi-batch design mitigates the coverage cost of representative-based clustering}
Because each cluster is inspected only through its centroid and a single outlier representative,
a cluster that mixes more than two root causes may leave a bug unexamined in one batch.
We do not actively filter such reports out of subsequent batches,
and missed bugs continue to surface as testing proceeds.
In our campaign,
a bug left unexamined in a Batch~\#1 cluster resurfaced as the outlier of a heterogeneous cluster in Batch~\#3 and was confirmed there,
providing an empirically viable trade-off between per-batch efficiency and overall coverage.

\paragraph*{Human validation shows that automated triage preserves real bugs}
To assess the correctness of the LLM analysis,
we sample 100 analyzed reports together with the conclusions produced by the multi-agent system
and validate them against human expert judgment.
\reftable{rq3-human-audit} summarizes the results.
Real-bug recall is 100\%, which means that all 29 sampled reports that correspond to real defects are retained after triage.
After root-cause deduplication, these 29 bug reports collapse to three unique bugs.
False-positive recall is 87.3\%, which shows that the LLM conservatively escalates some false-positive reports as potential bugs.
Escalating a false positive only adds review effort, whereas dropping a real bug is unrecoverable, so this asymmetry is a deliberate safety bias that lets the same run retain every real bug at $100\%$ recall.
The overall agreement rate between human reviewers and the LLM is 91\%.
Overall, although some wrong verdicts remain, the sampled audit shows that this error profile is acceptable for triage.

\ifdefined\TOSEMVERSION\else
\input{tables/05-rq3-human-audit}
\fi

\paragraph*{Issue-gate filtering removes duplicates rather than suppressing new bugs}
We also sample 100 reports filtered out by the issue gate
and check whether any of them corresponds to a new bug.
\reftable{rq3-gate-audit} shows that none of these reports corresponds to a new bug;
all of them are duplicates of previous issues.
Because the filter rules only narrow the scope of the triggering request,
the gate reduces duplicate review effort without suppressing new defects.

\ifdefined\TOSEMVERSION
\begin{table}[!htbp]
  \caption{Human audits of postprocessing decisions on NeuG:
  (a) reports analyzed by the multi-agent system and
  (b) reports filtered by the issue gate.}
  \label{tab:rq3-audits}
  \centering
  \begin{subtable}[t]{0.485\linewidth}
    \centering
    \caption{Analyzed reports.}
    \label{tab:rq3-human-audit}
    \scriptsize
    \setlength{\tabcolsep}{2.5pt}
    \renewcommand{\arraystretch}{1.06}
    \begin{tabular*}{\linewidth}{@{\extracolsep{\fill}}lrr@{}}
      \toprule
      \textbf{Metric} & \textbf{Count} & \textbf{Rate} \\
      \midrule
      Agreement with human verdict & 91 / 100 & 91\% \\
      Real-bug recall & 29 / 29 & 100\% \\
      False-positive recall & 62 / 71 & 87.3\% \\
      \bottomrule
    \end{tabular*}
  \end{subtable}
  \hfill
  \begin{subtable}[t]{0.485\linewidth}
    \centering
    \caption{Issue-gate exclusions.}
    \label{tab:rq3-gate-audit}
    \scriptsize
    \setlength{\tabcolsep}{2.5pt}
    \renewcommand{\arraystretch}{1.30}
    \begin{tabular*}{\linewidth}{@{\extracolsep{\fill}}lrr@{}}
      \toprule
      \textbf{Outcome} & \textbf{Count} & \textbf{Rate} \\
      \midrule
      Duplicate issue & 100 & 100\% \\
      New bug & 0 & 0\% \\
      \bottomrule
    \end{tabular*}
  \end{subtable}
\end{table}

\else
\input{tables/05-rq3-gate-audit}
\fi

\FloatBarrier
\subsection{RQ5: Bug Discovery in a Previously Untested System}
\label{sec:eval-bugdiscovery}

We apply the eight testing methods integrated in \sys to NeuG~\cite{NeuG},
a recently open-sourced graph database that has not previously been targeted by existing graph database testing methods,
to show the benefits of method integration in \sys.

\begin{figure}[!htbp]
  \centering
  \ifdefined\TOSEMVERSION
  \includegraphics[width=0.74\columnwidth]{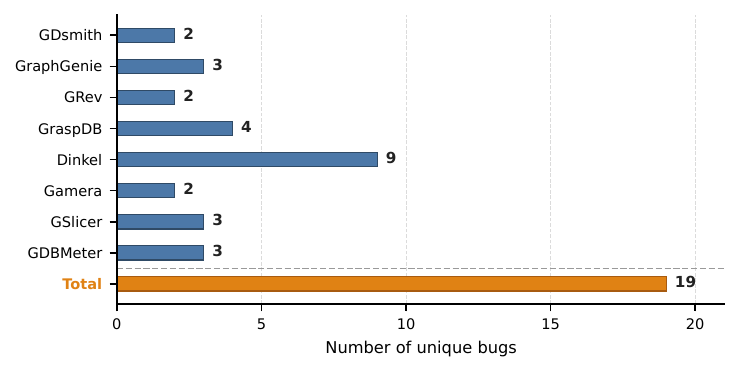}
  \else
  \includegraphics[width=\columnwidth]{figures/neug_bug_finders_bar.png}
  \fi
  \caption{Unique bugs uncovered on NeuG by each integrated method and in aggregate under \sys. A single bug may be discovered by multiple methods; per-method counts therefore do not sum to the total.}
  \Description{Bar chart of unique NeuG bugs uncovered per method: GDsmith 2, GraphGenie 3, GRev 2, GraspDB 4, Dinkel 9, Gamera 2, GSlicer 3, and GDBMeter 3, against 19 in aggregate under GraphWeave.}
  \label{fig:rq3-neug-bugs}
\end{figure}

\paragraph*{The unified platform saturates at 19 unique bugs, and no single method covers all identified bugs}
Each method, run in isolation, uncovers between 2 and 9 unique bugs on NeuG, as shown in \reffig{rq3-neug-bugs}.

We run each method until its unique-bug count stops growing, with new reports then filtered by the issue gate as duplicates of known issues.
These per-method counts therefore represent each method's discovery ceiling on this version rather than an artifact of insufficient runtime.
Because the per-method bug sets overlap, the aggregate is smaller than the sum of individual counts.
Combined under the unified platform, the eight methods collectively uncover 19 unique bugs, beyond which further runs surface only duplicates of known issues.
Dinkel is the most productive method, and its 9 bugs include almost every bug found by the other equivalence-based metamorphic methods,
yet Dinkel still misses the bugs found by methods such as GSlicer.
No single method finds all 19, so the methods are complementary rather than substitutable.

\paragraph*{The uncovered bugs are confirmed genuine defects}
Of the 19 unique bugs, 
all 19 have been reported upstream, all 19 have been confirmed as genuine defects, and 13 have already been fixed.

Together, the two RQs establish the practical value of \sys.
Its postprocessing workflow cuts the human-inspection workload to one sixth to one third of the raw reports while preserving every audited real-bug report, yielding $100\%$ real-bug recall; in the audit sample, the 29 retained bug reports collapse to three root-cause-level unique bugs after deduplication, closing the redundancy gap from the study without trading away coverage.
Additionally, its unified platform combines eight complementary methods to uncover 19 previously unknown bugs in a previously untested system.
A unified framework backed by process-level evidence thus delivers not only methodological value but also tangible returns in real-world bug discovery.

\section{Threats to Validity}
\label{sec:discussion}

\paragraph*{Internal validity}
Every bug count rests on human judgment.
We mitigate reviewer bias with the two-author protocol of \refsec{eval-setup}, and developers confirmed all 19 NeuG defects.
The automated triage of \refsec{whitebox} only aids reviewers and cannot silently drop a defect; \refsec{eval-workload} shows $100\%$ real-bug recall at $91\%$ human agreement.
For method fidelity, we port each method's construction logic and oracle unchanged, cross-validate each integration against its reference implementation, and count departures as method-level implementation-induced false positives (\reftable{fp-category-distribution}) rather than \sys failures.

\paragraph*{External validity}
We test capabilities from eight methods over four Cypher systems, with NeuG serving as the target outside prior graph database testing work, so results may not transfer to other query languages or architectures.
The two phenomena we report should nonetheless generalize.
Semantic-constraint false positives follow from standards that leave pattern matching semantics to implementations, and plan collapse follows from the optimization every mature engine performs.
The fixed 12-hour budget changes report volumes, not the existence claims behind developer-confirmed bugs.

\paragraph*{Construct validity}
We measure defects at the root-cause level after deduplication rather than by raw report volume, so repeated reports of one root cause do not inflate bug counts.
Plan collapse is measured by conservative within-system plan-hash comparison, so we can only undercount collapses.

\paragraph*{Conclusion validity}
Our analysis of false-positive causes in \refsec{study-rq2} is qualitative and does not aim to estimate their occurrence rates.
A single 12-hour campaign per method and system already yields 4{,}114 raw reports, far above the plausible number of real defects.
We identify these causes by inspecting reports and examining the underlying oracle assumptions and database behavior.
This qualitative analysis does not require repeated campaigns to estimate report frequencies.

\section{Related Work}
\label{sec:related}

\paragraph*{Graph and relational database testing}
Prior graph database testing contributes complementary oracles, including differential testing~\cite{hua2023gdsmith}, statement metamorphic methods~\cite{mang2024testing,jiang2024detecting,kamm2023testing,zhuang2023testing,liu2024testing,wust2024dinkel}, and graph-derived capabilities~\cite{zhuang2023testing,mang2025finding}.
These adapt ideas from relational DBMS testing, including differential testing~\cite{slutz1998massive}, random generation~\cite{SQLsmith}, and the SQLancer oracles~\cite{rigger2020testing,rigger2020detecting,rigger2020finding}.
Prior work primarily evaluates testing methods by the real bugs they find, with less attention to reports or tests that cannot reveal defects.
\sys integrates capabilities from eight methods behind a shared interface and examines both their reports and process-level data to study these problems.
Its optimizer-induced plan-collapse finding is a query-engine instance of the oracle-design challenges surveyed in~\cite{chen2018metamorphic}.

PATHTest~\cite{gui2026pathtest}, developed concurrently with our work, tests graph databases through transformations between fixed-length and variable-length queries.
Like our study, it reports and quantifies false positives across existing methods.
It attributes them to semantic differences and implementation errors.
Rather than propose another testing method, we examine why existing methods produce false positives.
We group them into four root-cause categories and focus on graph semantics.
We identify three aspects of pattern-matching semantics that affect test oracles.
Our criterion treats cross-system disagreement as a false positive when each system is internally consistent, and disagreement across equivalent forms within one system as a real bug.
Based on this analysis, we derive guidelines for oracle design and develop an LLM-based workflow for automatic report analysis.

\paragraph*{Portable testing and LLM-based testing}
Recent portable frameworks separate common logic from target handling through hand-maintained drop-ins~\cite{wust2024dinkel} or runtime feedback that reweights the generation space, as in SQLancer++~\cite{zhong2026scaling}.
These designs improve portability, but target knowledge still lives in the testing frontend as static rules or validity feedback that trades away parts of the generation space.
\sys instead derives target knowledge from post-run evidence and keeps dialect adaptation separate from report denoising, preserving testing intent by attributing each false positive to the assumption it violates under the underspecified GQL semantics~\cite{GQL2024,angles2017foundations}.
Whereas LLMs have been used to \emph{generate} tests~\cite{xia2024fuzz4all,yang2024whitefox}, \sys applies a multi-agent system to \emph{triage} their output under the evidence-backed acceptance criteria of \refsec{whitebox-multi-agent}.

\section{Conclusion}
\label{sec:conclusion}
In this paper, we present an empirical study of eight existing graph database testing methods through integration and process-level analysis, and distill guidelines for both integrated testing and future test-method design.

\paragraph*{For integrated testing}
False positives from query dialects and semantic constraints, together with duplicate reports, recur across methods, so a unified postprocessing pipeline is needed.
Within it, intra-system consistency serves as the principle that separates semantic false positives from true defects.
We realize these guidelines in \sys and confirm their actionability on NeuG~\cite{NeuG}, which also shows the value of integration.

\paragraph*{For future method design}
Semantic assumptions should be treated as explicit preconditions of oracle design.
In addition, equivalence-based metamorphic methods can benefit from skipping pairs that share a plan on the same graph state, which cannot expose defects, before execution.

\bibliographystyle{ACM-Reference-Format}
\bibliography{references}

\end{document}